\documentstyle[12pt,epsfig]{article}

\begin{document}
\input feynman
\renewcommand{\baselinestretch}{1.1}
\newcommand{\fig}[1]{Fig.\ref{#1}}
\newcommand{\tab}[1]{table \ref{#1}}
\newcommand{\figs}[2]{Figs.\ref{#1},\ref{#2}}
\newcommand{\tabs}[2]{tables \ref{#1}-\ref{#2}}
\newcommand{\amp}{{\cal A}}
\newcommand{\bq}{\begin{equation}}
\newcommand{\eq}{\end{equation}}
\newcommand{\bqa}{\begin{eqnarray}}
\newcommand{\eqa}{\end{eqnarray}}
\newcommand{\nl}{\nonumber \\}
\newcommand{\eqn}[1]{Eq.(\ref{#1})}
\newcommand{\eqns}[2]{Eq.(\ref{#1},\ref{#2})}
\newcommand{\suml}{\sum\limits}
\newcommand{\la}{\lambda}
\newcommand{\pr}{\partial}
\newcommand{\lmu}{_{\mu}}
\newcommand{\umu}{^{\mu}}
\renewcommand{\P}{{\cal P}}
\newcommand{\B}[1]{{\cal A}^{L}_{#1}}
\newcommand{\A}[1]{{\cal A}_{#1}}
\newcommand{\Ah}[2]{\hat{{\cal A}}_{#1#2}}
\def\demo{$\Delta\eta\mu \acute{o} \kappa \varrho \iota \tau o \varsigma$}
\pagestyle{empty}
\begin{flushright}
DEMO-HEP-96/03 \\
September 1996
\end{flushright}
 
\vspace*{2cm}
\begin{center}
\begin{huge}
{\bf Nullification in scalar theories\\[12pt] with derivative couplings}
\end{huge}
\\
\vspace{2\baselineskip}
{\bf\large Ernestos N.~Argyres$^{\dagger}$,  and 
Costas G.~Papadopoulos$^{\ddagger}$}\\
NRCPS \demo, Athens, Greece\\
\vspace{\baselineskip}
{\bf\large Maarten~Bruinsma$^{\star}$}\\
NIKHEF, Amsterdam, the Netherlands\\
\vspace{\baselineskip}
{\bf\large Ronald~Kleiss$^{\star\star}$}\\
University of Nijmegen, Nijmegen, the Netherlands\\
\vspace{5\baselineskip}
{\bf Abstract}\\[\baselineskip]
\end{center}
We discuss the structure of scalar field theories 
having the property that all {\it on-shell} $S$-matrix elements 
vanish in tree approximation. It is shown that there exists 
a large class of such theories, with derivative couplings,
which are all locally related to a free theory by a nonlinear transformation. 
It is also shown that a field-dependent wave-function renormalization
provides all necessary counterterms so that all on-shell $S$-matrix elements 
vanish also at the one-loop level. 

\vfill
\noindent\rule[0.in]{4.5in}{.01in} \\      
\vspace{.3cm}  
E-mail: 
\parbox[t]{12.6cm}{ {$^{\dagger}$argyres@cyclades.nrcps.ariadne-t.gr},
{$^{\ddagger}$Costas.Papadopoulos@cern.ch},
{$^{\star}$t26@nikhefh.nikhef.nl},{$^{\star\star}$kleiss@sci.kun.nl}.}

\newpage

\pagestyle{plain}
\setcounter{page}{1}

\section{Introduction}

\subsection{Threshold nullification in scalar theories}

The last few years, there has been considerable interest in the
high-multiplicity predictions of scalar field theories
\cite{general}. Originally motivated by considerations
on the possible enhancement of the rate of 
nonperturbative $B+L$-violating processes \cite{ringwald},
the study of multiparticle production in
perturbation theory has led to several new insights.
In the first place, it has been proven that,
at least in the tree approximation, the cross section
will violate the bounds imposed by unitarity when the
multiplicity and energy become sufficiently high
\cite{volo1,akp1}. Since the kinematics of many-particle
final states can become extremely involved, most 
exact results have been derived for situations where
all final-state particles are produced at rest. 
In addition, results have been derived for the case where
one or two final-state particles are off-shell. These
can be used in the evaluation of one- and two-loop
corrections \cite{akploop}. Another, and potentially
much more interesting, phenomenon is that of
{\em threshold nullification}  \cite{akp2,nullification}, which consists of the
following. In some scalar theories, the threshold
amplitudes $\amp(2\to n)$
 for the $2\to n$ process vanish identically
for an  infinite number of values of $n$. For instance,
a pure $\phi^3$ theory leads to $\amp(2\to n)=0$ for
all $n>3$; for a pure $\phi^4$ theory the same happens
for $n>4$, and in a spontaneously broken $\phi^4$ theory
even for $n>2$. Other, nonrenormalizable theories are
known showing related behaviour.
Apart from a relation between nullification 
and integrability of the classical action
\cite{integ}, no real physical reason for threshold 
nullification is evident. At any rate, above the kinematical
threshold such theories do not show exact nullification \cite{papa:bey-thr}, 
except for integrable theories in two dimensions, as for example
in  the sine-Gordon theory. 

\subsection{The Ooguri-Vafa vertex}
Several years ago, Ooguri and Vafa discussed the geometry
of $N=2$ strings, in which they discovered a similar nullification 
property also beyond the threshold
\cite{ooguri}. They illustrated this string theory by
a self-interacting scalar field theory defined on a space
with a $(1,1,-1,-1)$ metric. Let each momentum
$p$ be denoted by four real components, 
$(p^1,p^2,p^3,p^4)$, or, equivalently, by two
complex ones, $(p^1+ip^2,p^3+ip^4)$. The massless
scalar theory is then defined by the following three-point
interaction vertex:
$$
(p_1\cdot\bar{p}_2)(p_2\cdot\bar{p}_1)-
(p_1\cdot\bar{p}_1)(p_2\cdot\bar{p}_2)\;\;,
$$
where $p_{1,2}$ are two of the momenta incoming into the
vertex. It can be checked that, indeed, this theory has
the property that all tree-level $S$-matrix elements with
more than 3 external legs vanish identically
\cite{maarten}. In this
$2+2$-dimensional spacetime, a massless particle can
decay into two massless ones, so that the 3-point
$S$-matrix element is also physical. It is, however,
not invariant under general boosts: the symmetry of the
vertex is only $U(1,1)$ instead of $SO(2,2)$
\cite{berkowitz}, casting therefore some doubt
on the usefulness of this effective theory as a physically meaningful
one, even in 2+2 dimensions. A systematic search through a large
class of three-point vertices with four derivatives suggests
that no Lorentz-invariant
theory with `total nullification' can be constructed
from three-point interactions with four derivatives alone.

\section{Tree-order Amplitudes}

\subsection{The recursion relation}

\begin{figure}[ht]
\begin{center}
\fbox{
\begin{picture}(40000,16000)

\drawline\fermion[\E\REG](8000,8000)[5000]
\global\advance\pbackx by 2000
\put(\pbackx,\pbacky){\circle{4000}}
\global\advance\pbacky by -300
\global\advance\pbackx by -400
\put(\pbackx,\pbacky){$n$}
\global\advance\pbackx by  400
\global\advance\pbackx by 2500

\put(\pbackx,\pbacky){$=$}

\global\advance\pbackx by 2500
\global\advance\pbacky by 300
\drawline\fermion[\E\REG](\pbackx,\pbacky)[5000]
 
\drawline\fermion[\NE\REG](\pbackx,\pbacky)[5000]
\global\advance\pbackx by 1414
\global\advance\pbacky by 1414
\put(\pbackx,\pbacky){\circle{4000}}
\global\advance\pbacky by -300
\global\advance\pbackx by -400
\put(\pbackx,\pbacky){$n_1$}
\global\advance\pbackx by  400
 
\drawline\fermion[\E\REG](\pfrontx,\pfronty)[5000]
\global\advance\pbackx by 2000
\put(\pbackx,\pbacky){\circle{4000}}
\global\advance\pbacky by -300
\global\advance\pbackx by -400
\put(\pbackx,\pbacky){$n_2$}
\global\advance\pbackx by  400

\global\advance\pbacky by -2100
\global\advance\pbackx by -200
\put(\pbackx,\pbacky){\huge .}
\global\advance\pbackx by -300
\global\advance\pbacky by -500
\put(\pbackx,\pbacky){\huge .}
\global\advance\pbackx by -300
\global\advance\pbacky by -500
\put(\pbackx,\pbacky){\huge .}
\global\advance\pbackx by  1000
\global\advance\pbacky by  2900

\drawline\fermion[\SE\REG](\pfrontx,\pfronty)[5000]
\global\advance\pbackx by 1414
\global\advance\pbacky by -1414
\put(\pbackx,\pbacky){\circle{4000}}
\global\advance\pbacky by -300
\global\advance\pbackx by -400
\put(\pbackx,\pbacky){$n_p$}
\global\advance\pbackx by  400
 
\end{picture}
}
\caption[.]{Diagrammatic representation of
the recursion formula for the tree-order amplitudes.}
\label{fi01}
\end{center}
\end{figure}
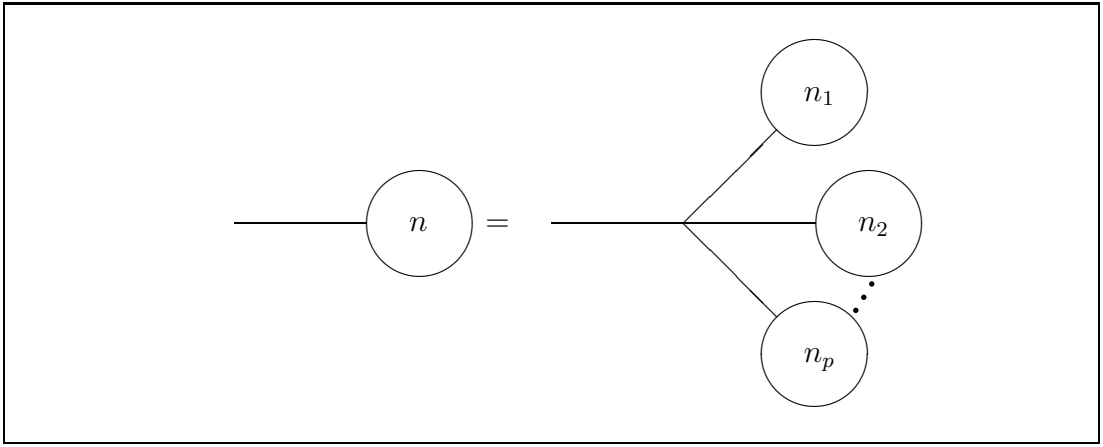

Motivated by the previous analysis we are studying
theories with two-deri\-va\-ti\-ve couplings.
In these theories the vertex, with $k$ external
scalar legs, is given by
\bq
-i\la_k\left(p_1^2+p_2^2+\cdots+p_k^2\right)-im^2\mu_k\;\;,
\eq
where $p_1,\ldots,p_k$ are the momenta in the legs, and 
$m$ is the mass of the scalar field. Notice that this is 
the most general Lorentz invariant and Bose symmetric vertex.
The dimensionality of both $\la_p$ and $\mu_p$ is
$2-p$. We shall allow for an arbitrary
number of such couplings, and write
\bqa
V(\phi) & = & \suml_{p\ge3}{\la_p\over p!}\phi^p\;\;,\nl
W(\phi) & = & \suml_{p\ge3}{\mu_p\over p!}\phi^p\;\;,\nl
\Omega(\phi) & = & {1\over2}m^2\phi^2 + m^2W(\phi)\;\;;
\eqa
The corresponding Lagrangian reads, of course,
\bq
{\cal L} = {1\over2}(\pr\lmu\phi)^2\left(1-2V^{(2)}(\phi)\right)
 - \Omega(\phi)\;\;.
\label{theory}
\eq
We shall compute the tree-level Green's functions with 
$(n+1)$ legs, $n$ of which have on-shell momenta $p_1,\ldots,p_n$,
with $p_1^2=\cdots=p_n^2=m^2$,
and the remaining one off-shell momentum is $p_0$, with $p_0^2=s$.
The on-shell legs are all truncated, but in the leg with momentum
$p_0$ the full propagator is kept. 
We denote the one-leg-untrancated amplitude by $\A{n}$. 
It is easy to see that $\A{0}=0$, and $\A{1}=1$.
The recusrion relation for $n>1$ reads, \fig{fi01},
\bqa
\A{[n]} & = & \suml_{p\ge2} {1\over p!(s-m^2)}
\suml_{[n_1],\ldots,[n_p]}
\A{[n_1]}\cdots \A{[n_p]}\nl
& & \hphantom{\suml_{p\ge2} {1\over p!(s-m^2)}}                 
\left(\la_{p+1}\left[s+s_{[n_1]}+\cdots+s_{[n_p]}\right]
+m^2\mu_{p+1}\right)\;\;,
\eqa
where 
\[
s_{[n]}=p^2_{[n]} \;\;\mbox{and}\;\; p^\mu_{[n]}=\sum_{i\in [n]} p^\mu_i\;.
\]
Here, we used the following notation. The set $[k_i]$
is a set of $k_i$ distinct labels taken from
$[n]$ = \{$1,2,3,\ldots,n$\}. In the recursion relation, 
the sets $[k_1], [k_2],\ldots, [k_p]$ do not intersect and
form a complete partition of $[n]$.

\subsection{Conditions on the potentials}

It is easily seen that, if the $\A{[k]}$ are pure numbers for $k<n$,
then $\A{[n]}$ will only depend on $s$ and $m^2$, and not on any
other dot products, because of the symmetry of the sum over
subsets:
\bqa
&&{\sum}^\prime \left(s+s_{[n_1]}+\cdots+s_{[n_p]}\right)= \nl
&&(s+n m^2)\frac{n!}{n_1!\cdots n_p!} + (s-n m^2)\left\{
\frac{(n-2)!}{(n_1-2)!\cdots n_p!}+\ldots+\frac{(n-2)!}{n_1!\cdots 
(n_p-2)!}\right\}\;,
\nl
\eqa
where the sum is performed over all subsets $[n_1],\ldots,[n_p]$ with
$n_1,\ldots,n_p$ fixed.
If in addition we manage to make $\A{n}$
also independent of $s$ and $m^2$, then the $S$-matrix elements will
vanish upon truncation of the remaining leg, and we have what we
call total nullification. We may therefore assume the $\A{[n]}$ to be all independent
of both $s$ and $m$, and examine to what requirements this leads
on the relation among the potentials $V(\phi)$ and $W(\phi)$. Therefore, we
put $\A{[n]} = \A{n}$, and find after some combinatorics
\bqa
(s-m^2)\A{n} & = & \suml_{p\ge3}{1\over p!}
\suml_{n_1,\ldots,n_p}\A{n_1}\cdots \A{n_p}\nl
& &                                                        
\left[(\la_{p+1}(s+nm^2)+m^2\mu_{p+1}){n!\over n_1!n_2!\cdots n_p!}
\right.\nl
& & \left.        
+\la_{p+1}(s-nm^2){p\;(n-2)!\over(n_1-2)!n_2!\cdots n_p!}\right]\;\;.
\eqa
As usual, we define a generating function as
\bq
f(z) = \suml_{n>0}\A{n}\;{z^n\over n!}\;\;,
\label{genfun}
\eq
so that $f(0)=0$ and $f'(0)=1$. Our recursion relation then
becomes the following differential equation, with
$\pr\equiv \pr/\pr z$:
\bqa
(s-m^2)f & = & (s-m^2)z + (s+m^2z\pr)V^{(1)}\nl
& & + (s-m^2z\pr)\pr^{-2}\left(f''V^{(2)}\right)
+ m^2W^{(1)}\;\;,
\label{bigeqn}
\eqa
with $\pr^{-2}$ defined by 
\[ \pr^{-2}F=G\;\; \Leftrightarrow \;\; G=\pr^{2}F \]
and 
\[ V^{(n)}=\frac{\pr^n V}{\pr f^n}\; .\]
Now, since the $\A{n}$ are by assumption independent of $s$ and $m^2$,
we may take $m=0$ in the above equation and apply $\pr^2$, to
end up with
\bq
f'' = 2f''V^{(2)}  + (f')^2V^{(3)} \;\;.
\eq
It is easily found that $f=f(z)$ and $z=\omega(f)$ must then
be related by
\bq
z = \omega(f) = \int\limits_{0}^{f}\;dx\;
\sqrt{1-2V^{(2)}(x)}\;\;.
\label{z=w}
\eq
Inserting this result in the $m^2$-dependent part of \eqn{bigeqn},
we then find that, in order to arrive at total nullification,
$W(f)$ must satisfy
\bq
W(f) = -{1\over2}f^2 + {1\over2}\omega(f)^2\;\;.
\label{W-potential}
\eq
Notice that in a massless theory total nullification exists
for arbitrary `potential' $V(\phi)$, whereas in the massive case the 
potential $W(\phi)$ is related to $V(\phi)$ through \eqn{W-potential}.
Moreover, if we rewrite our Lagrangian in terms of the
`composite' field $\omega(\phi)$ instead of $\phi$, we
obtain
\bq
{\cal L} = {1\over2}\left(\pr\lmu\omega\right)^2
-{1\over2}m^2\omega^2\;\;.
\label{free}
\eq
We see that, on the basis of this nonlinear transformation,
all scalar theories with total nullification at tree order
become locally equivalent to the free theory. It is also easy to verify that the
generating function is a solution of the classical field
equations \cite{brown}.

\section{One-loop Amplitudes}

\subsection{The one-loop recursion relation}

Although the theory for the field $\phi(x)$ as defined by \eqn{theory} 
seems to be related, at least locally, to
a free field theory of the `composite' field $\omega$, one might
expect that this is a classical result and at the one-loop level
the two theories will be very different. This reasoning is further supported by
the fact that the theory of \eqn{theory} is not renormalizable by the standard
power-counting criterion.

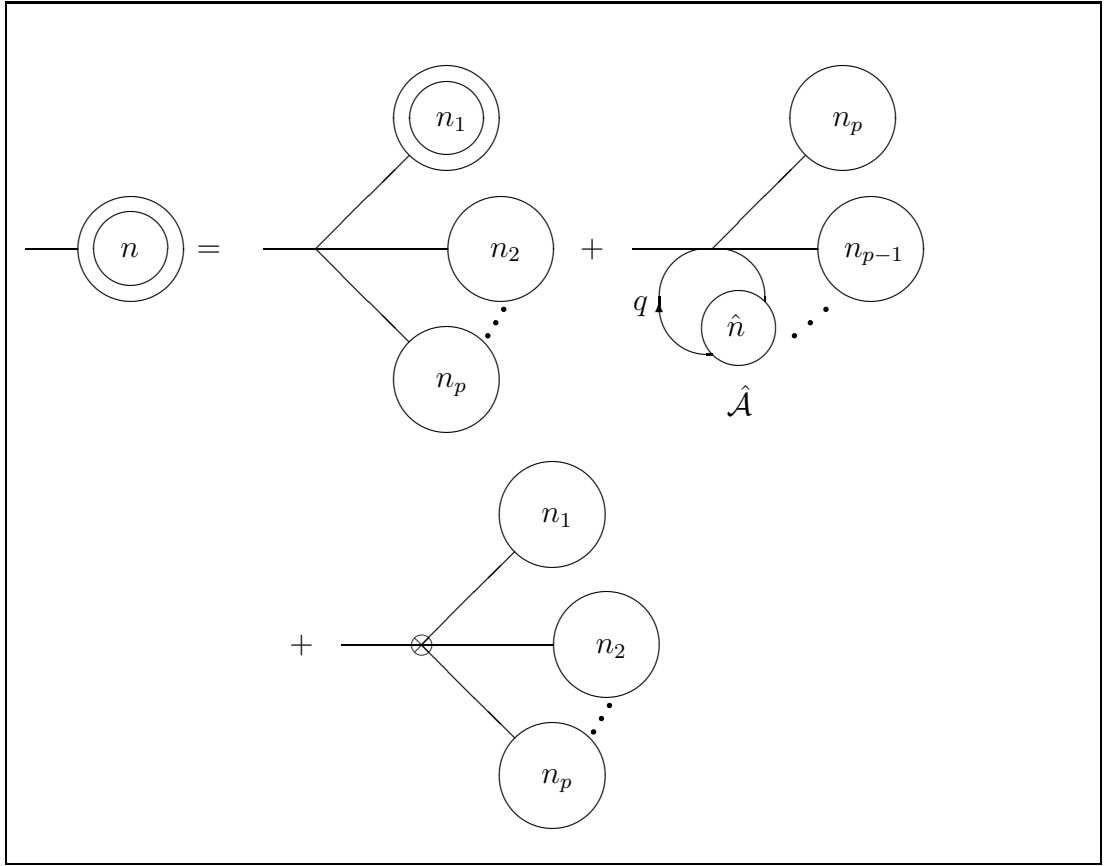
\begin{figure}[ht]
\begin{center}
\fbox{
\begin{picture}(40000,32000)
\drawline\fermion[\E\REG](0,23000)[2000]
\global\advance\pbackx by 2000
\put(\pbackx,\pbacky){\circle{4000}}
\put(\pbackx,\pbacky){\circle{3000}}
\global\advance\pbacky by -300
\global\advance\pbackx by -400
\put(\pbackx,\pbacky){$n$}
\global\advance\pbackx by  400
\global\advance\pbackx by 2500
\put(\pbackx,\pbacky){$=$}
\global\advance\pbackx by 2500
\global\advance\pbacky by 300
\drawline\fermion[\E\REG](\pbackx,\pbacky)[2000]
 
\drawline\fermion[\NE\REG](\pbackx,\pbacky)[5000]
\global\advance\pbackx by 1414
\global\advance\pbacky by 1414
\put(\pbackx,\pbacky){\circle{4000}}
\put(\pbackx,\pbacky){\circle{3000}}
\global\advance\pbacky by -300
\global\advance\pbackx by -400
\put(\pbackx,\pbacky){$n_1$}
\global\advance\pbackx by  400
 
\drawline\fermion[\E\REG](\pfrontx,\pfronty)[5000]
\global\advance\pbackx by 2000
\put(\pbackx,\pbacky){\circle{4000}}
\global\advance\pbacky by -300
\global\advance\pbackx by -400
\put(\pbackx,\pbacky){$n_2$}
\global\advance\pbackx by  400

\global\advance\pbacky by -2100
\global\advance\pbackx by -200
\put(\pbackx,\pbacky){\huge .}
\global\advance\pbackx by -300
\global\advance\pbacky by -500
\put(\pbackx,\pbacky){\huge .}
\global\advance\pbackx by -300
\global\advance\pbacky by -500
\put(\pbackx,\pbacky){\huge .}
\global\advance\pbackx by  1000
\global\advance\pbacky by  2900

\drawline\fermion[\SE\REG](\pfrontx,\pfronty)[5000]
\global\advance\pbackx by 1414
\global\advance\pbacky by -1414
\put(\pbackx,\pbacky){\circle{4000}}
\global\advance\pbacky by -300
\global\advance\pbackx by -400
\put(\pbackx,\pbacky){$n_p$}
\global\advance\pbackx by  400
 
\global\advance\pfrontx by 10000
\global\advance\pfronty by -300
 
\put(\pfrontx,\pfronty){$+$}
 
\global\advance\pfronty by 300
\global\advance\pfrontx by 2000

\drawline\fermion[\E\REG](\pfrontx,\pfronty)[3000]
 
\drawline\fermion[\E\REG](\pbackx,\pbacky)[4000]
\global\advance\pbackx by 2000
\put(\pbackx,\pbacky){\circle{4000}}
\global\advance\pbacky by -300
\global\advance\pbackx by -600
\global\advance\pbackx by -400
\put(\pbackx,\pbacky){$n_{p-1}$}
\global\advance\pbackx by  400
\global\advance\pbackx by  600

\global\advance\pbacky by -2100
\global\advance\pbackx by -2000
\put(\pbackx,\pbacky){\huge .}
\global\advance\pbackx by -600
\global\advance\pbacky by -500
\put(\pbackx,\pbacky){\huge .}
\global\advance\pbackx by -600
\global\advance\pbacky by -500
\put(\pbackx,\pbacky){\huge .}
\global\advance\pbackx by  3200
\global\advance\pbacky by  3100

\global\advance\pbackx by -6000
\global\advance\pbacky by  300

\drawline\fermion[\NE\REG](\pbackx,\pbacky)[5000]
\global\advance\pbackx by 1414
\global\advance\pbacky by 1414
\put(\pbackx,\pbacky){\circle{4000}}
\global\advance\pbacky by -300
\global\advance\pbackx by -400
\put(\pbackx,\pbacky){$n_{p}$}
\global\advance\pbackx by  400
 
\global\advance\pfronty by -2000
\put(\pfrontx,\pfronty){\oval(4000,4000)[tr]}
\put(\pfrontx,\pfronty){\oval(4000,4000)[l]}

\global\advance\pfrontx by -2000
\put(\pfrontx,\pfronty){\vector(0,1){0}}
\global\advance\pfrontx by -1000
\global\advance\pfronty by -300
\put(\pfrontx,\pfronty){$q$}
\global\advance\pfronty by  300
\global\advance\pfrontx by  3000

\global\advance\pfronty by -1000
\global\advance\pfrontx by 1000
\put(\pfrontx,\pfronty){\circle{2828}}
\global\advance\pfrontx by -450
\global\advance\pfronty by -300
\put(\pfrontx,\pfronty){$\hat{n}$}
\global\advance\pfronty by -3000
\put(\pfrontx,\pfronty){$\hat{\cal A}$}

 
 
\put(10000,7700){$+$}
 
\drawline\fermion[\E\REG](12000,8000)[3000]
 
\global\advance\pbackx by -450
\global\advance\pbacky by -300
\put(\pbackx,\pbacky){$\otimes$}
\global\advance\pbackx by 450
\global\advance\pbacky by 300
 
\drawline\fermion[\NE\REG](\pbackx,\pbacky)[5000]
\global\advance\pbackx by 1414
\global\advance\pbacky by 1414
\put(\pbackx,\pbacky){\circle{4000}}
\global\advance\pbacky by -300
\global\advance\pbackx by -400
\put(\pbackx,\pbacky){$n_1$}
\global\advance\pbackx by  400
 
\drawline\fermion[\E\REG](\pfrontx,\pfronty)[5000]
\global\advance\pbackx by 2000
\put(\pbackx,\pbacky){\circle{4000}}
\global\advance\pbacky by -300
\global\advance\pbackx by -400
\put(\pbackx,\pbacky){$n_2$}
\global\advance\pbackx by  400
 
\global\advance\pbacky by -2100
\global\advance\pbackx by -200
\put(\pbackx,\pbacky){\huge .}
\global\advance\pbackx by -300
\global\advance\pbacky by -500
\put(\pbackx,\pbacky){\huge .}
\global\advance\pbackx by -300
\global\advance\pbacky by -500
\put(\pbackx,\pbacky){\huge .}
\global\advance\pbackx by  1000
\global\advance\pbacky by  2900

\drawline\fermion[\SE\REG](\pfrontx,\pfronty)[5000]
\global\advance\pbackx by 1414
\global\advance\pbacky by -1414
\put(\pbackx,\pbacky){\circle{4000}}
\global\advance\pbacky by -300
\global\advance\pbackx by -400
\put(\pbackx,\pbacky){$n_p$}
\global\advance\pbackx by  400

\end{picture}
}

\caption[.]{Diagrammatic representation of
the recursion formula for the one-loop graphs (two circles).
The single circles are the tree-order amplitude. The blob with
two external legs corresponds to $\Ah{}{}$, as explained in the text.}
\label{fi02}
\end{center}
\end{figure}

Nevertheless, we have found that 
all one-loop divergencies of the one-leg-un\-trun\-ca\-ted amplitudes $\B{n}$,
are indeed canceled by counterterms which can be cast into a form 
identical to the bare lagrangian \eqn{theory},
using a non-linear wave-function renormalization. Let us be more
specific and write down the one-loop recursion relation corresponding to
\fig{fi02} :

\bq
(s-m^2)\B{[n]}=\P_1+\P_2 +\P_3
\label{1loop}
\eq
where $\P_1$ is the homogeneous term,
\bqa
\P_1&=&\sum \B{[n_1]}\A{n_2}\cdots\A{n_p}\frac{1}{(p-1)!}\nl 
&& \left[\left(
s+s_{[n_1]}+\ldots+s_{[n_p]}\right)\lambda_{p+1}+m^2\mu_{p+1}\right] \nl
\eqa
$\P_2$ is the inhomogeneous term,
\bqa
\P_2&=& \frac{1}{2}\sum\int \frac{d^Dq}{(2\pi)^D} (-i)\;\Ah{q}{[\hat{n}]} 
\A{n_3}\ldots\A{n_p}\frac{1}{(p-2)!} \nl
&&\left[\left(s+(q+p_{[\hat{n}]})^2+s_{[n_3]}+
\ldots+s_{[n_p]}\right)\lambda_{p+1}+m^2\mu_{p+1}\right] \nl
\eqa
and $\P_3$ is the counterterm contribution, to be determined,
\bqa
\P_3
&=& \sum \A{n_1}\A{n_2}\cdots\A{n_p}\frac{1}{p!}\nl 
&&\left[\left(
s+s_{[n_1]}+\ldots+s_{[n_p]}\right)\hat{\lambda}_{p+1}+m^2\hat{\mu}_{p+1}\right]
\nl
&+& \mbox{tadpole contributions.} \nl
\eqa
The one-loop counterterms can be written as
\bq
\hat{V}(\phi) = \suml_{p\ge3}{\hat{\la}_p\over p!}\phi^p\;\;\mbox{and}\;\;
\hat{W}(\phi) = \suml_{p\ge3}{\hat{\mu}_p\over p!}\phi^p\;.
\label{counterterms}
\eq

\subsection{Two-leg-untruncated amplitudes}

In order to proceed with \eqn{1loop} on has to calculate $\Ah{q}{[n]}$, which
is the amplitude with two untruncated external legs at tree order.
Since, by dimensional analysis, the dependence on the energy will be of the
form $E^{-2}$, we make the ansatz that 
\bq 
\Ah{q}{[n]} = \sum_{[k] \subset [n]\cup\{q\}} 
C_{[k]}\frac{i}{s_{[k]}-m^2}\;.
\eq

\begin{figure}[ht]
\begin{center}
\fbox{

\begin{picture}(40000,14000)

\drawline\fermion[\E\REG](7000,6000)[2000]
 
\global\advance\pbackx by 2000
\put(\pbackx,\pbacky){\circle{4000}}
\global\advance\pbackx by -450
\global\advance\pbacky by -300
\put(\pbackx,\pbacky){$n$}
 
\global\advance\pbackx by 2450
\global\advance\pbacky by  300
 
\drawline\fermion[\E\REG](\pbackx,\pbacky)[2000]
\put(\pbackx,\pbacky){\circle*{400}}
 
\global\advance\pbacky by  -1000
\put(\pbackx,\pbacky){$q$}
\global\advance\pbacky by   1000

\global\advance\pbackx by 2500
\global\advance\pbacky by -300
\put(\pbackx,\pbacky){$=$}
\global\advance\pbackx by 2500
\global\advance\pbacky by 300
 
\drawline\fermion[\E\REG](\pbackx,\pbacky)[2000]
 
\global\advance\pbackx by 2000
\put(\pbackx,\pbacky){\circle{4000}}
\global\advance\pbackx by -400
\put(\pbackx,\pbacky){$n_2$}
\global\advance\pbackx by  400
 
\global\advance\pbackx by 2000

\drawline\fermion[\E\REG](\pbackx,\pbacky)[3000]

\global\advance\pbackx by -1500
\put(\pbackx,\pbacky){\line(1,2){1500}}
\put(\pbackx,\pbacky){\line(-1,-2){1500}}
\global\advance\pbackx by  1500

\global\advance\pbackx by 2000
\put(\pbackx,\pbacky){\circle{4000}}
\global\advance\pbackx by -400
\put(\pbackx,\pbacky){$n_1$}
\global\advance\pbackx by  400

\global\advance\pbackx by 2000
\drawline\fermion[\E\REG](\pbackx,\pbacky)[2000]
\put(\pbackx,\pbacky){\circle*{400}}

\global\advance\pbacky by  -1000
\put(\pbackx,\pbacky){$q$}
\global\advance\pbacky by   1000

\end{picture}
}
\caption[.]{Diagrammatic representation of
the recursion relation for the amplitude with
two untruncated external legs.}
\label{fi03}
\end{center}
\end{figure}
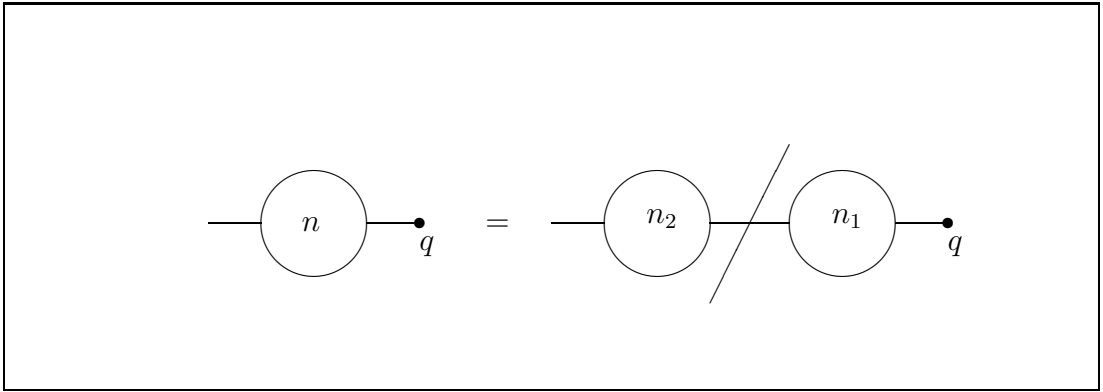
The next step is to consider the limit $s_{[k]}-m^2\to 0$ and try to identify
the residues of these simple poles. This can be done diagrammatically, by
cutting the graph as shown in \fig{fi03}. If the part of the
graph on the r.h.s. of the cut does not include the momentum $q$,
all legs of this subgraph are on-shell, and according to our previous results 
its contribution vanishes. 
The only non-vanishing contribution comes, 
from the r.h.s. subgraphs which include the momentum $q$, and
the result is simply given by:

\bq \Ah{q}{[n]} = \sum_{n_1+n_2=n} \A{n_1+1}\A{n_2+1}\sum_{[n_1]}\frac{i}
{\left(q+p_{[n_1]}\right)^2-m^2}\;.
\label{Ah}
\eq

\subsection{One-loop renormalization}

Using now \eqn{Ah} and performing the shift $q\to q-p_{[n_1]}$ the second term 
of the r.h.s of \eqn{1loop} is written as:

\bqa
\P_2&=& \frac{1}{2}\sum\int \frac{d^Dq}{(2\pi)^D} \frac{1}{q^2-m^2}
\A{n_1+1}\A{n_2+1}\A{n_3}\ldots\A{n_p}\frac{1}{(p-2)!} \nl
&&\left[\left(s+2q^2+s_{[n_1]}+
\ldots+s_{[n_p]}\right)\lambda_{p+1}+m^2\mu_{p+1}\right]\;. 
\eqa

It is now easy to see that in dimensional regularization only one form
of di\-ver\-gence\footnote{This is also true in
the Pauli-Villars regularization method.} appears, namely 
\[
{\cal I}=\int \frac{d^Dq}{(2\pi)^D} \frac{1}{q^2-m^2}
\]
so the previous equation can be cast into the form:
\bqa
\P_2&=& \frac{1}{2}{\cal I}\sum
\A{n_1+1}\A{n_2+1}\A{n_3}\ldots\A{n_p}\frac{1}{(p-2)!} \nl
&&\left[\left(s+2m^2+s_{[n_1]}+
\ldots+s_{[n_p]}\right)\lambda_{p+1}+m^2\mu_{p+1}\right]\;. 
\eqa

In order to calculate the counterterms needed to cancel the divergencies
of the one-loop amplitudes, we use the generating function technique  
by summing the inhomogeneous part, $\P_2+\P_3$ of 
\eqn{1loop} after multiplying by $z^n/n!$. This can be done because all
amplitudes $\A{}$ appearing in the inhomogeneous part are pure numbers, i.e.
they do not depend on any specific product $p_i \cdot p_j$. The result is 
\bqa 
\sum \left(\P_2+\P_3\right) \frac{z^n}{n!}&=&\frac{1}{2}{\cal I}\left\{
\left( s+2m^2+ m^2z\pr \right)f'^2 V^{(3)}\right.\nl
&+&\left. (s-m^2z\pr)\pr^{-2} \left[
2f'''f'V^{(3)}+f'^2f''V^{(4)}\right] 
+  m^2 f'^2 W^{(3)}\right\} \nl 
&+&(s+m^2z\pr)\hat{V}^{(1)}+(s-m^2z\pr)\pr^{-2}\left[f''\hat{V}^{(2)}\right]
+m^2 \hat{W}^{(1)}\nl
&+& \mbox{tadpole contributions}\;. \nl
\label{inhom-diff}
\eqa

The `renormalization' is now performed in the following sense:
starting with the bare lagrangian
\bq
{\cal L} = {1\over2}(\pr\lmu\phi_0)^2 \left(1-2V^{(2)}(\phi_0)\right)
- {1\over2} m^2 \phi_0^2 - m^2 W(\phi_0)\;\;,
\label{bare-l}
\eq
we define $\phi_0=Z^{1/2}\phi$, with $Z=1+{\cal I}^\prime h(\phi)$
where 
\bq
h(\phi)= \frac{f''}{f}+\frac{f'}{f}\left(b-2cz\right)\;.
\label{Z-h}
\eq
Notice that the r.h.s. of \eqn{Z-h} is expressed as 
a function of $z$, but $z$ should be understood as a function
of $\phi$, given by $z = \omega[\phi]$  (see \eqn{z=w}).
Moreover the infinite subtraction constant ${\cal I}^\prime$ is such that 
${\cal I}^\prime-{\cal I}=\mbox{finite}$.
This finite difference between
${\cal I}^\prime$ and  ${\cal I}$ is of course attributed to the
subtraction scheme. In order to avoid any unnecessary, for the moment,
complication let us choose ${\cal I}^\prime={\cal I}$, which corresponds to 
a total subtraction scheme: this is possible since only one form of
divergence, proportional to ${\cal I}$, appears in the \eqn{inhom-diff}.
Then, expanding \eqn{bare-l} and keeping linear terms in ${\cal I}$, we obtain
\bq
{\cal L} = {1\over2}(\pr\lmu\phi)^2 
\left(1-2V^{(2)}(\phi)-2\hat{V}^{(2)}(\phi)\right)
- {1\over2} m^2 \phi^2 - m^2 W(\phi) - m^2 \hat{W}(\phi)\;\;,
\eq 
where
\bq\hat{V}^{(2)}={\cal I}\left(
\frac{1}{2} \frac{f''^2}{f'^4}-\frac{1}{2} \frac{f'''}{f'^3}
+c\frac{1}{f'^2}\right)
\label{vhat}
\eq
and 
\bq \hat{W}={\cal I}\left( \frac{b}{2} z-c z^2+
\frac{z}{2}\frac{f''}{f'}\right) \;. 
\label{what}
\eq
The dimensionful constants $b$ and $c$ are determined by the requirement
that $\hat{V}$ and $\hat{W}$, start with $\phi^3$.

After some algebraic effort\footnote{We have verified the results using
\verb+Mathematica+} one can see that the expression \eqn{inhom-diff} 
becomes free of divergencies, even more becomes identically null, if the
`tadpole' contributions are given by
\[
\frac{b}{2}m^2{\cal I}\frac{1}{f'} \]
which is nothing but the `tadpole' counterterm of the free theory
\bq 
\delta{\cal L}=\frac{b}{2} m^2 {\cal I} \omega.
\eq
Having succeeded in canceling the divergencies, one can easily
see from \eqn{1loop} that $\B{n}=0$: this is because
the recursion relation is linear with respect to $\B{n}$ and moreover 
$\B{0}=\B{1}=0$. Notice that, in the total subtraction
scheme, there is no mass renormalization at the one-loop level. 

Of course in general the renormalization of this theory, even at the one-loop approximation,
is actually much more involved and we are not going to further continue the
discussion in this paper. For instance one might consider renormalization
of some coupling constants introduced by the potentials $V(\phi)$ and $W(\phi)$, 
so that knowledge of their specific form is necessary.
Nevertheless 
our results suggest that, although the theory defined by \eqn{theory}
is not power-counting renormalizable, divergencies of the one-leg-un\-trun\-ca\-ted 
amplitudes can be canceled by appropriate counterterms, 
as we have explicitly proven at the one-loop level. Moreover these
counterterms can be cast into
a form that is identical to the bare lagrangian, and hence equivalent to the 
free $\omega$-field theory, by using a field-dependent wave-function 
renormalization as well as the `tadpole' counterterm of the free theory.
What actually happens is the following: since the theory is not 
power-counting renormalizable
divergencies of a generic Green's function are not canceled by the counterterms
given in \eqns{vhat}{what}. Nevertheless
their combination corresponding to the one-leg-un\-trun\-ca\-ted
amplitude are free of divergencies. In that sense, one-leg-untruncated 
amplitudes as well as the on-shell $S$-matrix elements
are ultraviolet finite quantities. Moreover, in a total subtraction scheme,
they are all vanishing!
This total nullification seems to be related to the existence
of the nonlinear transformation $\phi_0 \to \omega[\phi_0]$, which
transforms the bare interacting lagrangian into the free one. 
It is not inconceivable that the equivalence between the two formulations, i.e.
in $\phi$ and $\omega$, as far as the $S$-matrix elements are concerned,
may be extended to all orders in perturbation theory.

\section{Discussion}

In order to get some more insight on the physics issues,
let us recall the example of a 
$\Phi^4$ theory with spontaneously-broken reflection symmetry,
which after the appropriate shift, $\Phi=\phi+v$, with $v=\langle\Phi\rangle$, 
takes the form:
\bq
 {\cal L}={1\over2}(\pr\lmu\phi)^2
 - {1\over2}m^2\phi^2\left(1+\frac{\phi}{2v}\right)^2\;.
\eq
The learned reader will recall that all
$\amp(2\to n)$ threshold amplitudes vanish for $n\ge 3$. 
It is indeed surprising that the new theory 
resulting by simply adding a three- and four-point derivative couplings, namely 
\bq
 {\cal L}={1\over2}(\pr\lmu\phi)^2\left(1+\frac{\phi}{v}\right)^2
 - {1\over2}m^2\phi^2\left(1+\frac{\phi}{2v}\right)^2\;,
\eq
exhibits the nullification property for all on-shell\footnote{We would
like to emphasize that nullification now extends over all physical phase-space.}
$S$-matrix elements !
The one-leg-untruncated amplitudes $\A{n}$ are given by ($n\ge 2$)
\bq
 \A{n}=(-v)^{1-n} (2n-3)!!\;.
\eq
In fact the `composite' field $\omega$ is given in this case by
a simple polynomial relation:
\bq
 \omega=\phi\left(1+\frac{\phi}{2v}\right)
\eq
and has the same physical mass, $m$, as the elementary field.
Quantum corrections make the above relation look more complicated,
but still the theory possesses the total-nullification  property.

\begin{figure}[htb]
\begin{center}
\mbox{\psfig{file=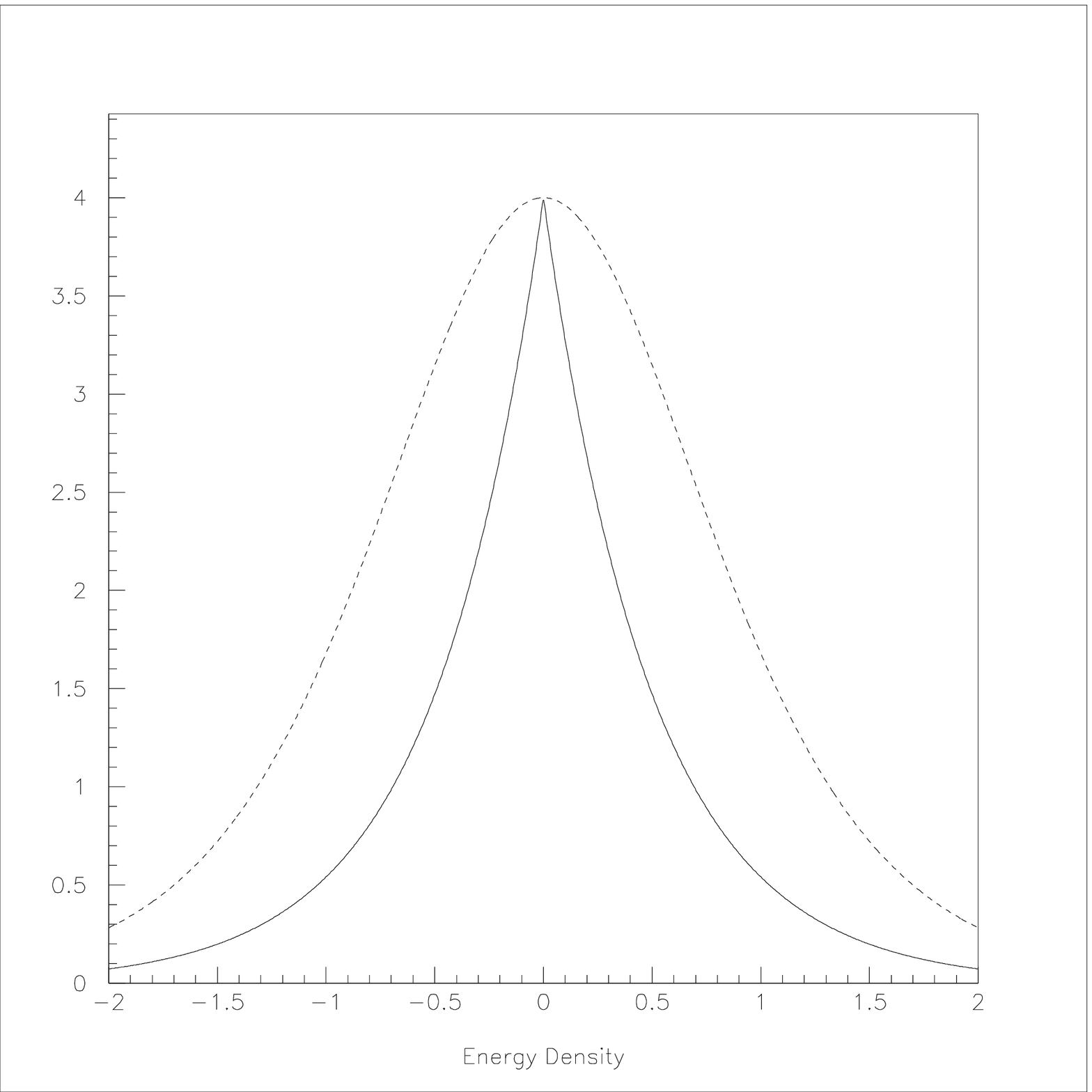,width=15cm,height=15cm}}
\caption[.]{The energy density of the solution $\phi(x)$ given in \eqn{soli}.
The dotted line represents the energy density of the sine-Gordon soliton-solution.}
\label{fig1}
\end{center}
\end{figure}

Another interesting example is that of the sine-Gordon theory. The
original lagrangian is given by 
\bq 
{\cal L}=\frac{1}{2}\left( \pr_\mu \phi\right)^2-\frac{m^2}{\lambda^2} 
\left( 1-\cos(\lambda\phi)\right)\;.
\eq
The $\omega$-field is now defined by
\bq
\omega[\phi]=\frac{2}{\lambda} \sin\frac{\lambda\phi}{2} 
\label{f-to-omega}\eq
and the new lagrangian is written as
\bq 
{\cal L}=\frac{1}{2}\left( \pr_\mu \phi\right)^2 \cos^2\left(\frac{\lambda\phi}{2}\right)
- \frac{m^2}{\lambda^2} \left( 1-\cos(\lambda\phi)\right)\;.
\label{new-sine-Gordon}\eq
Looking for finite energy stationary-solutions in 1+1 dimensions
we find that
\bq
\phi(x)=\frac{1}{\lambda}\left\{ \begin{array}{ll}
-2 \sin^{-1}\left( e^{-m(x-x_0)} \right) +2 \pi & x\ge x_0 \\
2 \sin^{-1}\left( e^{m(x-x_0)} \right)  & x\le x_0 \;. 
\end{array} \right.
\label{soli} 
\eq
The energy or the `mass' of this solution is given by
\bq
M_\phi= \frac{4 m}{\lambda^2}\;.
\eq
Moreover one can easily check that the virial theorem is indeed satisfied.
Notice that the sine-Gordon soliton is given by
\bq
\phi(x)=\frac{4}{\lambda} \tan^{-1}( e^{m(x-x_0)} )
\eq
and that the corresponding `mass' is
\bq
M_{sG}=  \frac{8 m}{\lambda^2}\;.
\eq
It is easy to see that both solutions, satisfy $\phi(-\infty)=0$ 
and $\phi(\infty)=2\pi/\lambda$.
In \fig{fig1} we show the energy-density profile of our solution, as well as
that of the sine-Gordon soliton, for $m=1$, $\lambda=1$ and $x_0=0$.

What is the moral of this sine-Gordon-like example ? The answer is more or less
obvious: the new theory defined by \eqn{new-sine-Gordon}, although it is locally 
equivalent to a free theory, by means of the transformation \eqn{f-to-omega},
it possesses new solutions, which do not exist in the free theory: it is well known that
the only finite-energy solution of the free-theory is $\omega(x)=0$.
This of course might have been suspected by the simple remark that the transformation 
\eqn{f-to-omega} implies some kind of a global constraint on the field-magnitude;
for instance if we restrict ourselves to real solutions then 
$|\omega(x)|\le 2/\lambda, \; \forall\; x$. Therefore theories with derivative
couplings such as the ones studied in this paper are not
trivially equivalent to the free theory.

Let us summarize our results by sketching the physical picture 
emerging from them. 
It is possible to consider scalar fields $\phi(x)$ with very complicated 
and highly non-trivial self-interactions described by \eqn{theory}. 
Nevertheless their apparent
dynamics is such that it has no measurable effects: all $S$-matrix elements 
are vanishing (as we have explicitly proven up to the one-loop level).  
This total nullification is equivalent to the
existence of a `composite' field $\omega[\phi]$, whose dynamics, at least
in peturbation theory,  is described by a locally free field theory:
{\it the $\omega$ field possesses no self-interactions}.

\vspace*{1cm}
\noindent {\Large\bf Acknowledgements} \\[12pt]
C.G.P. would like to thank the Department of Physics of the 
Catholic University of Nijmegen, where part of this work was done, 
for its kind hospitality. This work was partially supported by the EU 
grant CHRX-CT93-0319.

\end{document}